# THE APPLICATION OF CAUSE EFFECT GRAPH FOR THE COLLEGE PLACEMENT PROCESS

[1]Mrs.Dhanamma Jagli, [2]Mrs.Mamatha T, [3]Ms.Swetha Mahalingam, [4]Ms.Namrata Ojha

[1]Assistant Professor,
Department of MCA, V.E.S. Institute of Technology, Mumbai, India
dhana1210@yahoo.com

[2]Assistant Professor,
Department of CSE, SNIST, Hyderabad, India
mamathat7@gmail.com

[3,4]MCA Final year Students
V.E.S. Institute of Technology, Mumbai, India
[3]swethamahalingam@yahoo.com,
[4]namrataojha05@gmail.com

## ABSTRACT

*This paper presents a case study on the application of cause effect graph for representing the college placement process. This paper begins with giving a brief overview of the college placement process which will serve as the basis for developing the cause effect graph and the decision table for the same in a systematic manner. Finally, it concludes with the design of test cases thus giving a complete and clear representation about the application of cause-effect graph in the software testing domain.*

## 1. INTRODUCTION

The placement process in the college is a yearly affair brimming with activities. It is a process involving the active participation of the placement cell and the final year students of all streams. The process starts with the placement cell giving the students necessary information concerning the placement process. This dissemination of information requires to be done in a very concise and systematic manner and this is where comes in the application of cause effect graph.

The following are the main reasons for presenting this technical paper:

- ✓ For representing the information concerning the placement process in a clear and systematic manner to all the students and cater to different levels of understanding.
- ✓ This paper will serve as a valuable addition to the existing repository of case studies based on cause effect graph.
- ✓ To highlight the application of cause effect graph as an effective testing technique.

## 2. LITERATURE REVIEW

The Software testing technique Cause-Effect Graph was made-up by Bill Elmendorf of IBM in 1973[1] [2]. Instead of the test case designer trying to manually determine the right set of test cases, modeled the problem using a cause-effect graph, and the software that supports the technique. Cause Effect Graph creates a relation between the outputs (conditions) and the inputs (actions).A cause effect graph in software testing is an intended for graph that maps a number of conditions to a number of actions.

The techniques in the past introduced regarding the different input data as self-determining, and the input value each one considered independently for producing test cases. The different inputs and their effects on the outputs are not clearly measured for test case design [1] [2] [5].

In precedent only some case studies were at hand wherever Cause effect graph application had taken consideration. Such are Withdrawal at ATM, Purchase order of Different equipment, Employee payroll system, some are with designed test cases and some are with test cases [3]. But in this paper a new case study had chosen with outcome as test cases for automated college placement cell as shown in further stages in this paper.

## 3. OUTLINE OF CAUSE EFFECT GRAPH

A Boolean graph reflecting logical relationships between inputs (causes), and the outputs (effects) or transformations (effects).Cause-effect graphing describes a technique that uses the dependencies for identification of the test cases known as cause-effect graphing. The logical associations between the conditions (Causes) and their actions (effects) in a constituent or a system are represented is called as a cause-effect graph.

### a) Notations for Cause-and-Effect Graph

From the requirement specification, the possible causes (conditions) and effects (actions) must be identified. Each condition is described as a cause that consists of input conditions (or combinations of those) [1] [3]. The causes are associated with logical operators (e.g., AND, OR and NOT). A condition or cause can be correct or incorrect. The basic notations for causes-effects relations and its related constraint symbols were shown in the below Fig 1 and Fig 2.The Causes and effects are treated in the same way and noted in the graph as shown in the below Fig 3.

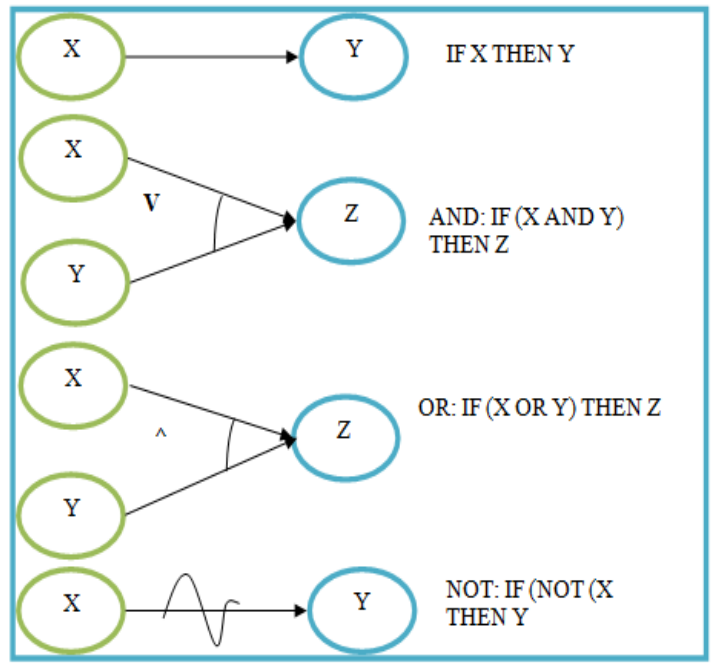

Figure 1: Basic Notations

Additional Notations for Cause-and-Effect Graph are different constraints that would be used in cause effect graph for showing association between Causes as well as effects Additional notations are shown in the below Fig 2.

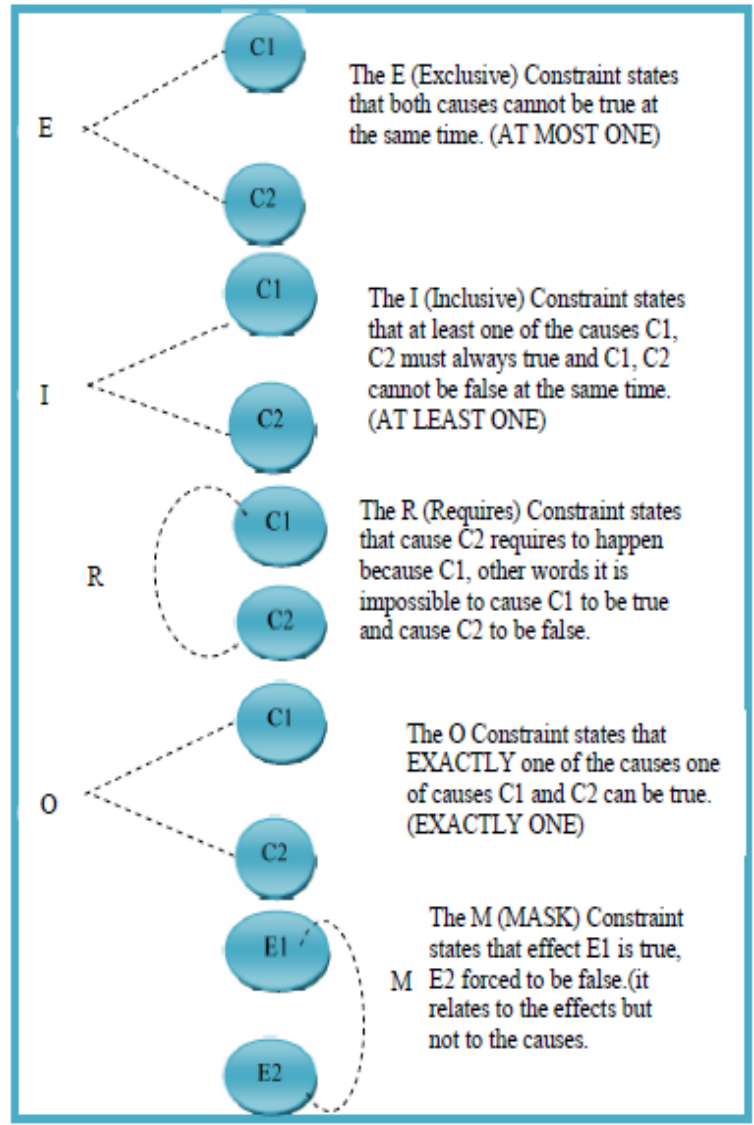

Figure 1: Basic Constraints Symbols

### b) Procedure to generate test cases

Test technique based on cause effect graph consists in the following steps:

1. Identify the causes and the effects
2. Establish the graph of relations between causes and effects
3. Complete the graph by adding constraints between causes and effects
4. Convert the graph to a decision table
5. Generate a test cases

In order to design proper set of test cases cause effect graph is not sufficient, so that it has to convert in to decision table.

### c) Cause Effect Graph to Decision Table

The Cause Effect Graph must be converted d into a decision table from that the test cases can be Considered. The steps to transform a graph into a table are as follows:

1) Choose an effect
2) Looking in the graph, find combinations of causes that have this effect and combinations that do not have this effect
3) Add one column into the table for every one of these cause-combinations and the caused states of the remaining effects
4) Ensure if decision table values occurred many times and, if yes, remove them.

The test cases based on decision tables had the intention to design test cases for executing attractive combinations of causes that possible failures can be found, in addition the causes and effects, transitional results may be added in the decision table.

A decision table had two parts. The first half (upper half), describes about the inputs (causes). The second half (lower half) describes about the outputs (effects).each column is a test case that the combination of causes or inputs and the anticipated effects or outputs for this association. Test cases intended for optimal situation, every combination of causes or inputs is considered as a test case [3] still, causes may influence or exclude each other in such a way that not all combinations construct valid test case. The implementation of every cause or condition and effect or action is represented with a "YES or 1" or "NO or 0". Each cause or condition and effect or action must occur at least once with either "YES" or "NO" in the table.

An optimized decision table does not include all possible combinations, but the impracticable combinations are not added. Based on the dependency between the inputs (causes) and the results (actions), the optimized decision table represents the result every column of decision table must be compiled as a test case.

### d) Definition of the Test Exit Criteria

Simple criteria for test completion, criteria for test completion can be defined relatively easily.

Executing every column in the decision table as a test case [3] is the minimum requirement of decision table based test cases in the software testing, that all possible combinations of conditions (causes) and corresponding effects (actions) can be verified.

The value of the technique is organized and very formal approach in constructing a decision table with all possible combinations may expose association between set of causes and set of effects that are not included when using other techniques for designing test cases.

### e) Benefits of Using a Cause-and-Effect

✓ Helps determine root causes.
✓ Encourages group participation.
✓ Uses an orderly, easy-to-read format.
✓ Indicates possible causes of variation.
✓ Increases process knowledge.
✓ Identifies areas for collecting data.

# 4. PROBLEM SPECIFICATION

Generate decision table based test cases for the placement process in a college open to engineering students of all streams and M.C.A students.

Definition of the various terminologies in use:

- ✓ A dream company is one that offers a package of Rs.5, 00,000/- or more per annum.
- ✓ A non-dream company is one that offers a package less than Rs.5, 00,000/- per annum.
- ✓ A core company is one that pertains to one's branch of study. For example: IT companies are considered to be core companies for M.C.A, B.E (computer-science), and B.E (I.T).
- ✓ A non-core company is one that does not pertain to one's branch of study.
- ✓ Some companies only offer placements and some others offer only internship and might absorb students eventually based on their performance.

The following are the set of requirements (constraints):

1. If a student is placed in a dream company then, he/she is out of the placement process.
2. If a student gets a $2^{nd}$ offer then the $1^{st}$ offer stands cancelled and he/she is also out of the placement process.
3. A student has only maximum 3 attempts for $2^{nd}$ job.
4. If a student gets placed in a core company then he/she can apply only for core companies for the $2^{nd}$ job.
5. If a student gets dream internship he/she is eligible for only non-dream companies for the $2^{nd}$ job.

# 6. ANALYSIS AND DESIGN

*STEP 1:* Various causes and effects had been identified to design the cause effect graph from the given problem statement.

The various causes identified from the given problem statement are as follows:

        C1:$1^{st}$ job is a dream, core/non-core job.
        C2:$1^{st}$ job is a non-dream, core job
        C3:$1^{st}$ job is a non-dream, non-core job
        C4:$1^{st}$ job is a dream, core/non-core internship
        C5:$1^{st}$ job is a non-dream, core/non-core internship
        C6:$2^{nd}$ job offer secured

The various effects identified from the given problem statement are as follows:

        E1: Out of the placement process
        E2:$1^{st}$ job offer stands cancelled
        E3: Maximum 3 attempts for only core, dream/non-dream job
        E4: Maximum 3 attempts for core/non-core, dream/non-dream job
        E5: Maximum 3 attempts for core/non-core, non-dream job.

STEP 2: at this moment, with the identified causes and effects, the cause-effect graph had been constructed.
*STEP 3:* Completed the graph by adding all identified constraints as shown in below Fig 3.

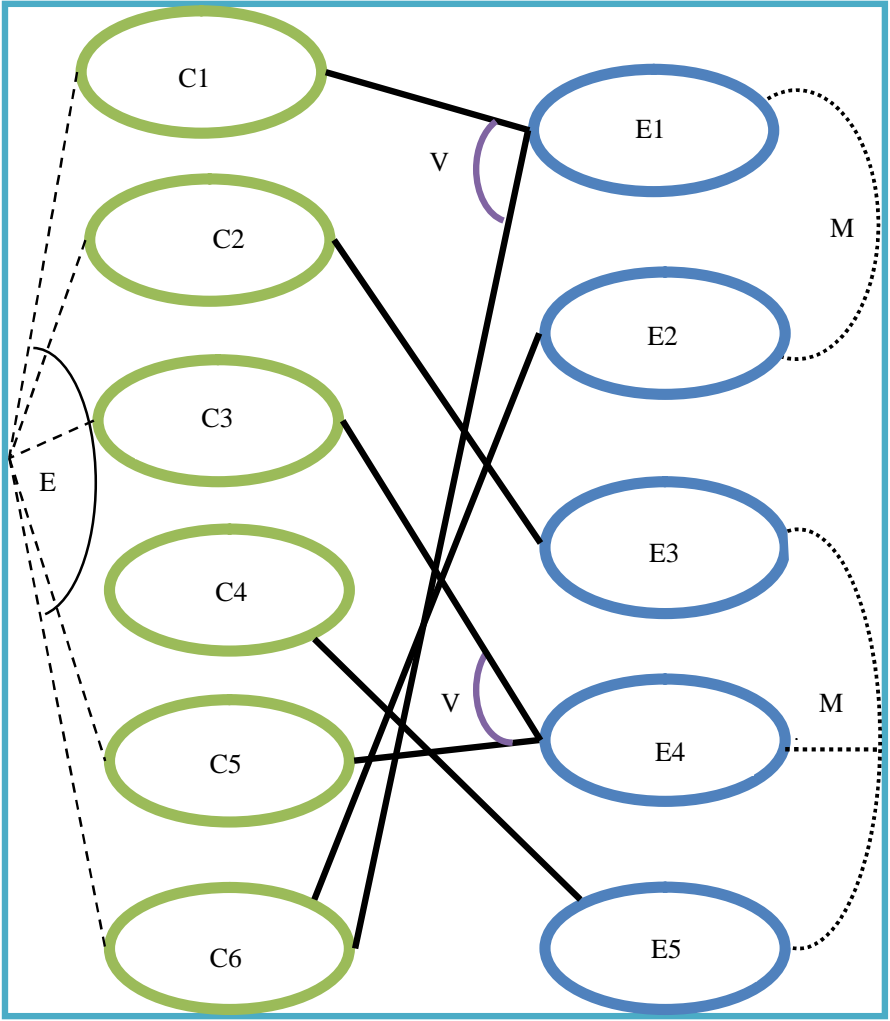

Figure 2: Cause Effect Graph for Placement cell

*STEP 4:* Constructed Cause-Effect Graph had transformed into a decision table as shown in below Table 1.

Table 1: Decision Table

| CONDITION/CAUSE | 1 | 2 | 3 | 4 | 5 |
|---|---|---|---|---|---|
| C1 | YES | - | - | - | - |
| C2 | - | - | YES | - | - |
| C3 | - | - | - | YES | - |
| C4 | - | - | - | - | YES |
| C5 | - | - | - | YES | - |
| C6 | YES | YES | - | - | - |
| EFFECT /ACTION | | | | | |
| E1 | YES | - | - | - | - |
| E2 | - | YES | - | - | - |
| E3 | - | - | YES | - | - |
| E4 | - | - | - | YES | - |
| E5 | - | - | - | - | YES |

*STEP 5:* Generated test cases for the automated college placement cell problem statement. Since there are 5 effects there should be at least 5 test cases.

The following are the test cases for the given problem statement as shown in following Table 2.

Table 2: Test cases

| Test Case | Input(causes) | Expected Output (Effects) |
|---|---|---|
| 1 | 1st Job Dream   OR   2nd Job secured | Out of the placement process |
| 2 | 2nd Job secured | 1st Job Cancelled |
| 3 | 1st Job Non-Dream, Core | Maximum 3 attempts for Core, Dream/Non-Dream |
| 4 | 1st Job Non-Dream, Non-Core OR 1st Job Non-Dream, Core/Non-Core Internship | Maximum 3 attempts for Core/Non-Core, Dream/Non-Dream |
| 5 | 1st Job Dream, Core/Non-Core Internship | Maximum 3 attempts for Core/Non-Core, Non-Dream |

# 7. CONCLUSION

It ought to be important that the case study used in this paper to illustrate the basic steps of Cause Effect Graph was kept very short and snappy. An experienced software tester could probably jump right to this set of test cases from the requirements without using the CEG method for small projects but  for large, complex systems with multiple causes (inputs) and effects (outputs or transformations) this method is an organized way to analyze the problem statement and create test cases[7]. The case study described in this paper would be helpful to the college students to understand the placement procedure easily. Apart from this, it would be useful to software testers, students, faculties and others interested in the software testing domain.

## ACKNOWLEDGEMENTS

As authors we would like to articulate thankfulness to Prof.A.Nagananda, Training and placement officer in our college for providing complete information about placement process takes place in the college.

## Authors

### Mrs.Dhanamma Jagli

Short Biography:

Mrs.Dhanamma Jagli is an Assistance professor in V.E.S Institute of Technology, Mumbai, currently Pursuing Ph.D in Computer Science and Engineering and received M.Tech in Information Technology from Jawaharlal Nehru Technological University, Hyderabad and Andhra Pradesh. She has around 9 years teaching experience at the post graduate and under graduate level. She had published and presented papers in referred international journals and conferences. Her areas of research interest are Data Mining, Cloud Computing, Software Engineering, Data base Systems and Embedded Real time systems. She has been associated with Indian Society of Technical Education (ISTE) as a life member.

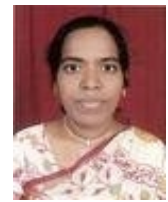

**Mrs.Mamatha T**

Short Biography:

Mrs. T.Mamatha, Working as an Assistant Professor in Dept of C.S.E of SreeNidhi Institute of Science &Technology. M.Tech in Software Engineering from JNTU, Anantapur, B.Tech in C.S.E from JNTUH.Her areas of research include Software Engineering, Network Security.

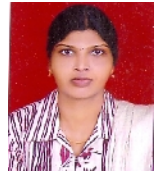

**Ms.Swetha Mahalingam**

Short Biography:

Swetha Mahalingam is a final year student of Master of Computer Application (M.C.A) from Vivekanand Education Society's Institute of Technology (V.E.S.I.T), Mumbai University.Swetha has completed her B.Sc. in Computer-science from Ramnarain Ruia College; Mumbai University. She is an ardent programmer with an abiding interest in software engineering and programming languages like C, C++, Java and .Net technologies.

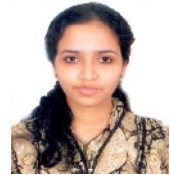

**Ms.Namrata Ojha**

Short Biography:

Namrata Ojha completed her BSc. degree in Information Technology from Sathaye College, Mumbai in the year 2010. She is currently pursuing MCA degree from Vivekanand Education Society's Institute of Technology, Mumbai.She has been selected as an Associate Software Engineer at Accenture Services Pvt Ltd.She has given Technical Paper Presenatation on the topic 'Autonomic Computing' at S.I.E.S. College's Inter-collegiate I.T. Fest and won 2nd prize for the same.Her areas of interests include Programming, Network Security and Software Engineering.

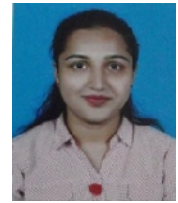